# Distributed Optimal Frequency Control for Integrated Energy Systems with Electricity and Heat

Xin Qin, Xuan Zhang, Xinwei Shen*, Yinliang Xu
Tsinghua-Berkeley Shenzhen Institute
Tsinghua University
Shenzhen, China
sxw.tbsi@sz.tsinghua.edu.cn

Mohammad Shahidehpour,
*Fellow*, IEEE
the Robert W. Galvin Center for Electricity Innovation
Illinois Institute of Technology
Chicago, USA

Hongbin Sun,
*Fellow*, IEEE
Department of Electrical Engineering
Tsinghua University
Beijing, China

*Abstract*—With more and more distributed energy resources (DERs) deployed in Integrated Energy Systems (IESs), frequency stability challenges the pursuit of reliability and efficiency. This paper proposes a fully-distributed frequency control method for load-side DERs, in which the optimality can be guaranteed in an IES where electricity and heat are coupled. Moreover, the global asymptotic stability of the closed-loop system is proved and the robustness with respect to inaccurate coefficients is shown. Case studies demonstrate the effectiveness of proposed method.

*Index Terms*--Integrated Energy System (IES), distributed frequency control, combined heat and power (CHP)

## I. Introduction

Integrated Energy Systems (IESs), also known as Multi-Energy Systems or multi-energy carriers, can bring high-efficient and environmentally-friendly multiple energy supply in distribution or transmission level. For example, British has an energy plan called "Thousands of Flowers" which supplies local energy with lots of distributed energy resources (DERs) such as combined heat and power (CHP) units and photovoltaic units [1]. However, when increasing DERs with uncertain renewables and electric-heat coupling CHP units integrated in an IES, the power mismatch between generation-side and load-side will lead to frequency deviation and threaten the reliability of energy supply. Thus, it is essential to propose a control method to stabilize the frequency and optimize the system economic efficiency as well for the IES.

Traditionally, frequency control is a hierarchical control with a centralized control center [2][3]. But this manner faces challenges in the IES because of expensive ramping reserves, heavy communication burden among a large number of DERs and slow reaction since the time scales of each layer are different. Currently load-side frequency control becomes popular due to its benefits, e.g. low adjustment cost, light communication burden and fast response.

In the electric power system (EPS), researchers have proposed load-side frequency management by controlling some types of loads or devices. For example, flexible load demand was utilized in [4] and [5] to respond to frequency regulation signals. Moreover, electric vehicles [6] and battery storage systems [7] were also controlled to provide ancillary services and realize load-side frequency management. These works illustrated how to model and control specific types of load-side devices, however, the global optimality of EPS's economic efficiency might not be guaranteed at the same time. To address this issue, [8] proposed a distributed proportional-integral (PI) load controller. [9] and [10] developed an optimal distributed control scheme to regulate frequency and voltage. Nevertheless, some reliability constrains such as load power and line power flow limits are not considered. Based on reverse engineering, [11]-[13] designed a distributed load-side control scheme to restore frequency with optimality, line thermal limit and convergence guaranteed under inaccurate coefficient, where [13] realized the fully distributed algorithm with only neighborhood communication and did not need accurate measurement of parameters. However, above methods face challenges in the IES because load-side electric power could depend on its heat demand. For example, if we directly apply the EPS frequency control in the IES, the electric power adjustment may break the operating constraints on CHP units, which can threaten the stability of frequency.

In an IES with electricity and heat, the district heat system (DHS) has flexibility but constraints on frequency regulations. For example, [14] realized frequency regulation by adjusting water boiler, [15] used heating, ventilation, and air-conditioning devices to support frequency regulation, and [16] adopted CHP units for microgrid frequency management. However, global optimality is not considered above.

In this paper, we propose a distributed load-side frequency control method for DERs with system-wide optimality ensured in the transmission level of IES, which is applied in secondary control. To be more specific, physically, our EPS frequency regulation considers constraints and flexibility from the DHS, and power load limits and line flow limits are also included; mathematically, detailed models of load-side DER and heat system are considered in the optimal control problem compared with [11]-[13]. It is also noteworthy that our method 1) only needs neighborhood and local information and is a fully distributed method, and 2) is robust to inaccurate coefficients and globally asymptotically stable.

## II. OPTIMAL FREQUENCY CONTROL MODEL FOR THE IES

### A. Preliminaries

We describe the IES topology as a directed graph $G(N,\varepsilon)$, in which $N=\{1,\cdots,n\}$ is the set of buses, and $\varepsilon = N \times N$ is the set of single-direction power lines. The buses generate or consume electricity and heat, while the power lines only transmit electricity. The buses are divided into two sets: generator bus set $G$ and load bus set $L$ with $N = G \cup L$. The generator buses contain generators and may contain attached loads, but the load buses only contain loads. We define $C \in |N| \times |\varepsilon|$ as the incidence matrix of graph $G(N,\varepsilon)$ where $C_{i,l}=1$ if $l=ij \in \varepsilon$, $C_{i,l}=-1$ if $l=ji \in \varepsilon$, and $C_{i,l}=0$ otherwise.

### B. Network Model

Using the DC power flow equation, line power flow $P_{ij}$ is calculated by:

$$P_{ij} = B_{ij}(\theta_i - \theta_j) \qquad (1)$$

where $B_{ij}$ is a constant. $\theta_i$ and $\theta_j$ indicate the phase angle of bus $i$ and $j$. For simplicity, all variables such as $P_{ij}, \theta_i, \theta_j$ are defined as the deviations from their nominal values $P_{ij}^0, \theta_i^0, \theta_j^0$ calculated by economic dispatch.

Consider the dynamic network model of an IES shown in Fig.1 with electricity and heat:

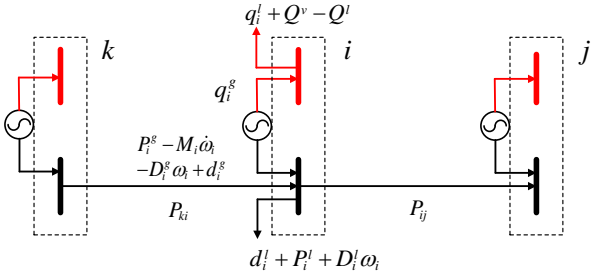

Figure 1. Dynamics of frequency at bus $i$

where $\omega_i$ is the frequency of bus $i$, $d_i$ and $q_i$ are the controllable electric and heat loads, in which $d_i = d_i^l - d_i^g$ and $q_i = q_i^l - q_i^g$. $M_i$ is the generator inertia, and the aggregated uncontrollable power injection $P_i^{in}$ is the difference of the uncontrollable generator power output $P_i^g$ and the load power $P_i^l$, i.e. $P_i^{in} = P_i^g - P_i^l$, and the uncontrollable heat power injection $Q_i^{in}$ is the uncontrollable load power $-Q_i^l$, i.e. $Q_i^{in} = -Q_i^l$. Damping coefficient $D_i$ is the sum of generator and load damping coefficients denoted by $D_i^g$ and $D_i^l$ respectively, i.e. $D_i = D_i^g + D_i^l$. In minutes or seconds, heat loads have heat inertia, thus:

$$Q_i^{in} - q_i = Q_i^v \qquad (2)$$

where $Q_i^v$ is the heat inertia limited by the lower boundary $\underline{Q}_i^v$ and the upper boundary $\overline{Q}_i^v$. By introducing $Q_i^v$, small mismatch of heat power can be compensated by its heat inertia, which increases the flexibility of the IES.

Electricity and heat are coupled via generators which are modeled as CHP units whose feasible region is a convex combination of extreme points. Here we use $q_i \le k_m d_i + b_m$, $\forall m \in K_+$ and $q_i \ge k_n d_i + b_n, \forall n \in K_-$ to describe the feasible region, where $k_i$ and $b_i$ are coefficients, and $K_+$ and $K_-$ are sets of upper and lower boundaries, respectively.

Defining $P^e = CP$, where $P=\{P_{ij}\}_{i \in N}$, as the leaving line power, network dynamics can be modeled by:

$$M_i \dot{\omega}_i = P_i^{in} - d_i - D_i \omega_i - P_i^e \quad \forall i \in G \qquad (3)$$
$$0 = P_i^{in} - d_i - D_i \omega_i - P_i^e \quad \forall i \in L \qquad (4)$$
$$\dot{P}_{ij} = B_{ij}(\omega_i - \omega_j) \quad \forall ij \in \varepsilon \qquad (5)$$

where (3) and (4) represent dynamics of generator buses and load buses. (5) reflects line power flow dynamics under the assumption that the frequency deviation is small.

**Remark 1:** Practically, the uncontrollable electric power injection and damping coefficient $P_i^{in}$ and $D_i$ are hard to measure accurately. These errors may lead to frequency instability. Our proposed method does not need the measurement of $P_i^{in}$, and is robust to inaccurate $D_i$, which will be strictly proved in Theorem 2.

### C. Optimal Load-Side Frequency Control Problem

The steady state of the IES is provided by economic dispatch, i.e. system (3)-(5) are adopted around an equilibrium with $\omega_i = 0 \ \forall i \in N$ and $\dot{P}_{ij} = 0 \ \forall ij \in \varepsilon$. If the disturbance reflected by any step change of $P_i^{in}$ or $Q_i^{in}$ occurs in the IES, our goal is to 1) restore frequency to the nominal value i.e. 50Hz or 60Hz; 2) rebalance system electric and heat power and let each control area absorb its power imbalance; 3) achieve minimal cost under IES operating constrains.

To realize above goals, following optimal load-side frequency control (OLFC) problem is proposed:

$$\min_{\omega,d,q,P,\varphi} f = \sum_{i \in N}[C_{i,e}(d_i) + C_{i,h}(q_i)] + \sum_{i \in N}\frac{1}{2}D_i\omega_i^2 \qquad (6a)$$

s.t.
$$P_i^{in} - d_i - D_i\omega_i - \sum_{j:ij\in\varepsilon} P_{ij} + \sum_{k:ki\in\varepsilon} P_{ki} = 0 \qquad (6b)$$

$$P_i^{in} - d_i - D_i\omega_i - \sum_{j:ij\in\varepsilon_{in}} B_{ij}(\varphi_i - \varphi_j) + \sum_{k:ki\in\varepsilon_{in}} B_{ki}(\varphi_k - \varphi_i) = 0 \qquad (6c)$$

$$q_i \le k_m d_i + b_m, \forall m \in K_+ \qquad (6d)$$
$$q_i \ge k_n d_i + b_n, \forall n \in K_- \qquad (6e)$$
$$\underline{d}_i \le d_i \le \overline{d}_i \qquad (6f)$$
$$\underline{Q}_i^v \le q_i - Q_i^{in} \le \overline{Q}_i^v \qquad (6g)$$
$$\underline{P}_{ij} \le B_{ij}(\varphi_i - \varphi_j) \le \overline{P}_{ij} \qquad (6h)$$

where $\varphi_i$ is the virtual phase angle [12] to eliminate the measurement of real phase angle $\theta_i$. (6b)-(6h) are for $\forall i \in N$. (6h) is for $\forall ij \in \varepsilon$. (6a) is the objective function which minimizes the cost. (6b) is a redundant equation for algorithm design, and (6c) ensures that power imbalance is absorbed in each control area. (6d) and (6e) are electric and heat power limits of CHP units. (6f)-(6h) are the limits of electric power, heat power, and phase angle. For optimality and convergence analysis, we have the following assumptions:

**Assumption 1:** $C_{i,e}(d_i)$ and $C_{i,h}(q_i)$ are strictly convex and continuously differentiable with $C_{i,e}''(d_i) \ge \alpha > 0$.

**Assumption 2:** The Problem (6) is feasible.

## III. THE DISTRIBUTED OPTIMAL FREQUENCY CONTROL ALGORITHM

In this section, a fully-distributed algorithm is proposed to solve the OLFC problem (6) based on the reverse engineering [11]-[13]. The derivation process includes three steps: 1) derive Lagrangian function; 2) apply partial primal-dual gradient method derive a distributed control scheme; 3) propose the implementation framework of the scheme.

### A. Lagrangian Function Derivation

Firstly, we derive the Lagrangian function from (6):

$$L = \sum_{i \in N} C_{i,e}(d_i) + C_{i,h}(q_i) + \sum_{i \in N} \frac{1}{2} D_i \omega_i^2$$
$$+ \sum_{i \in N} \lambda_i (P_i^{in} - d_i - D_i \omega_i - \sum_{j:ij \in \varepsilon} P_{ij} + \sum_{k:ki \in \varepsilon} P_{ki})$$
$$+ \sum_{i \in N} \mu_i [P_i^{in} - d_i - \sum_{j:ij \in \varepsilon_{in}} B_{ij}(\varphi_i - \varphi_j) + \sum_{k:ki \in \varepsilon_{in}} B_{ki}(\varphi_k - \varphi_i)]$$
$$+ \sum_{m \in K^+} \sum_{i \in N} \zeta_i^m (q_i - k_m d_i - b_m) - \sum_{n \in K^-} \sum_{i \in N} \zeta_i^n (q_i - k_n d_i - b_n)$$
$$+ \sum_{i \in N} \gamma_i^+ (d_i - \underline{d_i}) - \sum_{i \in N} \gamma_i^- (d_i - \overline{d_i})$$
$$+ \sum_{i \in N} \delta_i^+ (-q_i + Q_i^{in} - \overline{Q}_i^v) + \sum_{i \in N} \delta_i^- (q_i - Q_i^{in} + \underline{Q}_i^v)$$
$$+ \sum_{i \in N} \sigma_{ij}^+ [B_{ij}(\varphi_i - \varphi_j) - \underline{P}_{ij}] - \sum_{ij \in \varepsilon} \sigma_{ij}^- [B_{ij}(\varphi_i - \varphi_j) - \overline{P}_{ij}]$$

### B. Partial Primal-dual Gradient Method Application

Secondly, a partial primal-dual gradient method is applied to reduce the number of variables and derive the control mechanism of OLFC problem. Define $\alpha = (\gamma, \delta, \sigma)$ and reduce the variable $\omega$ by defining:

$$\overline{L}(d,q,P,\varphi,\lambda,\mu,\zeta,\alpha) = \min_{\omega} L(\omega,d,q,P,\varphi,\lambda,\mu,\zeta,\alpha) \quad (7)$$

then eliminate variable $\lambda_L$ by defining:

$$\hat{L}(d,q,P,\varphi,\lambda_g,\mu,\zeta,\alpha) = \min_{\lambda_L} \overline{L}(d,q,P,\varphi,\lambda,\mu,\zeta,\alpha) \quad (8)$$

Notice that (9) needs the measurement of the aggregated power injection $P_i^{in}$:

$$\dot{\mu}_i = \varepsilon_{\mu_i}[P_i^{in} - d_i - \sum_{j:ij \in \varepsilon_{in}} B_{ij}(\varphi_i - \varphi_j) + \sum_{k:ki \in \varepsilon_{in}} B_{ki}(\varphi_k - \varphi_i)] \quad (9)$$

thus a new variable $r_i$ is introduced to eliminate $P_i^{in}$ in the algorithm where $r_i = \frac{K_i}{\varepsilon_{\mu_i}} \mu_i - \frac{K_i}{\varepsilon_{\lambda_i}} \omega_i$.

Other variables in the partial primal-dual gradient method are:

$$\dot{d}_i = -\varepsilon_{d_i}(C'_{i,e}(d_i) - \frac{\varepsilon_{\lambda_i}}{\varepsilon_{\lambda_i} + \varepsilon_{\mu_i}}\omega_i - \frac{\varepsilon_{\mu_i}}{K_i} r_i + \gamma_i^+ - \gamma_i^-$$
$$- \sum_{m \in K^+} \zeta_i^m k_m + \sum_{n \in K^-} \zeta_i^n k_n) \quad (10a)$$

$$\dot{q}_i = -\varepsilon_{q_i}(C'_{i,h}(q_i) - \delta_i^+ + \delta_i^- + \sum_{m \in K^+} \zeta_i^m - \sum_{n \in K^-} \zeta_i^n) \quad (10b)$$

$$\dot{P}_{ij} = \varepsilon_{P_{ij}}(\omega_i - \omega_j) \quad (10c)$$

$$\dot{\varphi}_i = -\varepsilon_{\varphi_i}[\sum_{j:ij \in \varepsilon_{in}} B_{ij}(\mu_i - \mu_j - \sigma_{ij}^+ + \sigma_{ij}^-)$$
$$- \sum_{k:ki \in \varepsilon_{in}} B_{ki}(\mu_k - \mu_i - \sigma_{ki}^+ + \sigma_{ki}^-)] \quad (10d)$$

$$\dot{r}_i = K_i[D_i \omega_i + \sum_{j:ij \in \varepsilon} P_{ij} - \sum_{k:ki \in \varepsilon} P_{ki} - \sum_{j:ij \in \varepsilon_{in}} B_{ij}(\varphi_i - \varphi_j)$$
$$+ \sum_{k:ki \in \varepsilon_{in}} B_{ki}(\varphi_k - \varphi_i)] \quad (10e)$$

$$\dot{\zeta}_i^n = \varepsilon_{\zeta_i^n}[q_i - k_n d_i - b_n]_{\zeta_i^n}^+ \quad (10f)$$

$$\dot{\zeta}_i^m = \varepsilon_{\zeta_i^m}[-q_i + k_m d_i + b_m]_{\zeta_i^m}^+ \quad (10g)$$

$$\dot{\gamma}_i^+ = \varepsilon_{\gamma_i^+}[d_i - \overline{d}_i]_{\gamma_i^+}^+ \quad (10h)$$

$$\dot{\gamma}_i^- = \varepsilon_{\gamma_i^-}[-d_i + \underline{d}_i]_{\gamma_i^-}^+ \quad (10i)$$

$$\dot{\delta}_i^+ = \varepsilon_{\delta_i^+}[Q_i^{in} - q_i - \overline{Q}_i^v]_{\delta_i^+}^+ \quad (10j)$$

$$\dot{\delta}_i^- = \varepsilon_{\delta_i^-}[-Q_i^{in} + q_i + \underline{Q}_i^v]_{\delta_i^-}^+ \quad (10k)$$

$$\dot{\sigma}_{ij}^+ = \varepsilon_{\sigma_{ij}^+}[B_{ij}(\varphi_i - \varphi_j) - \underline{P}_{ij}]_{\sigma_{ij}^+}^+ \quad (10l)$$

$$\dot{\sigma}_{ij}^- = \varepsilon_{\sigma_{ij}^-}[-B_{ij}(\varphi_i - \varphi_j) + \overline{P}_{ij}]_{\sigma_{ij}^-}^+ \quad (10m)$$

where (10a) and (10b) are for $i \in G$, (10c) and (10l)-(10m) are for $ij \in \varepsilon$, and (10d)-(10k) are for $i \in N$. The operator $[w]_v^+$ indicates if $w > 0$ or $v > 0$, $[w]_v^+ = w$, otherwise $[w]_v^+ = 0$, so $[w]_v^+ \leq w$ [12]. $\mu_i$ in (10d) is calculated from $r_i$ and $\omega_i$.

**Theorem 1 (Global asymptotic convergence):** Under Assumptions 1 and 2, the algorithm (10) with the network model (3)(4) converge to the optimal point $(\omega^*, d^*, q^*, P^*, \varphi^*, \lambda^*, \mu^*, \varepsilon^*, \gamma^*, \zeta^*, \alpha^*)$ asymptotically where $(\omega^*, d^*, q^*, P^*, \varphi^*)$ is the optimal solution of problem (6).

### C. Algorithm Implementation

Thirdly, the implementation of (10) is shown in Fig. 2:

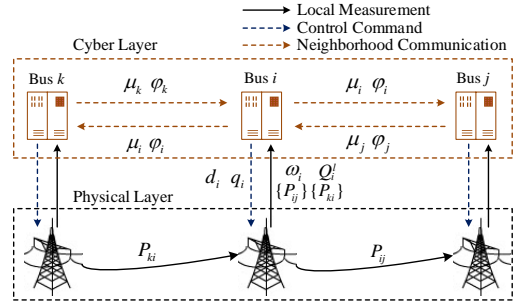

Figure 2. The implementation of the distributed control algorithm (10)

The implementation includes two steps: information gathering and control demand executing. In the physical layer, each bus measures its local frequency deviation and line power flow deviations of EPS, and updates the local heat load power deviation which is influenced by heat demand, outer temperature, etc. In the cyber layer, each bus exchanges $\mu_i(t)$ and $\varphi_i(t)$ with adjacent buses. Then, each bus computes its control variables $d_i(t+1)$ and $q_i(t+1)$ and sends them to the physical layer to adjust power. The loop of frequency response and control command generation is closed to restore frequency with optimality guaranteed.

**Theorem 2 (Convergence with inaccurate coefficients):** If Assumption 2 works with following assumptions held:

1) $C_{i,e}(d_i)$ and $C_{i,q}(q_i)$ are $\alpha$ strongly convex and second-order continuously differentiable, and $C'_{i,e}(d_i)$ and $C'_{i,h}(q_i)$ are Lipschitz continuous for a Lipschitz constant $L > 0$

2) The coefficient $\varepsilon_{d_i}$ of (10a) and $\varepsilon_{q_i}$ of (10b) are large enough so that following conditions are satisfied:

$$C'_{i,e}(d_i) - \frac{\varepsilon_{\lambda_i}}{\varepsilon_{\lambda_i} + \varepsilon_{\mu_i}}\omega_i - \frac{\varepsilon_{\mu_i}}{K_i}r_i + \gamma_i^+ - \gamma_i^- - \sum_{m \in K^+}\zeta_i^m k_m + \sum_{n \in K^-}\zeta_i^n k_n = 0$$

$$C'_{i,h}(q_i) - \delta_i^+ + \delta_i^- + \sum_{m \in K^+}\zeta_i^m - \sum_{n \in K^-}\zeta_i^n = 0$$

3) Define the inaccurate damping coefficient $\tilde{D}_i = D_i + \tau a_i$, where the inaccurate coefficient $\tau a_i$ satisfies:

$$\tau a_i \in 2(\underline{d}' - \sqrt{\underline{d}'^2 + \underline{d}'D_{\min}}, \underline{d}' + \sqrt{\underline{d}'^2 + \underline{d}'D_{\min}})$$

where $\underline{d}' = 1/L$ and $D_{\min} = \min_{i \in N} D_i$.

The closed-loop system (3), (4), and (10) converges to a point $(\omega^*, d^*, q^*, P^*, \varphi^*, \lambda^*, \mu^*, \varepsilon^*, \gamma^*, \zeta^*, \alpha^*)$ where $(\omega^*, d^*, q^*, P^*, \varphi^*)$ is the optimal solution of problem (6), even under inaccurate information of coefficients.

## IV. PROOF OF OPTIMALITY AND CONVERGENCE

### A. Proof of Theorem 1

Define $v = (d, q)$, $x = (\varphi, P)$, and $y = (\mu_L, \mu_g, \lambda_g, \zeta, \alpha)$. The control mechanism can be written as:

$$\dot{d} = -\Gamma_d\frac{\partial \hat{L}}{\partial d} \quad \dot{q} = -\Gamma_q\frac{\partial \hat{L}}{\partial q} \quad \dot{x} = -\Gamma_x\frac{\partial \hat{L}}{\partial x} \quad \dot{y} = -\Gamma_y\left[\frac{\partial \hat{L}}{\partial y}\right]_y^+ \quad (11)$$

where $\Gamma_d = diag(\varepsilon_d)$ is the diagonal matrix of the positive coefficient of step sizes, etc. Define $o = (d, q, x)$, and $z^* = (o^*, y^*)$ to be any equilibrium of (10). Give:

$$U_{z^*}(z) = \frac{1}{2}(z - z^*)^T \Gamma_z^{-1}(z - z^*) \quad (12)$$

where $\Gamma_z$ is a block diagonal matrix consisting of corresponding entries $(\Gamma_o, \Gamma_y)$.

According to Assumption 1, $L$ is strictly convex in $d$ and $q$. In addition, it can be proved that $L$ is strictly concave in $\lambda_g$ and linear in other variables. As a result, $L(o^*, y) - L(o^*, y^*) \leq 0$, and $L(o^*, y^*) - L(o, y^*) \leq 0$. Thus:

$$\dot{U}_{z^*}(z) = -(o - o^*)^T\frac{\partial L}{\partial o} + (y - y^*)^T\left[\frac{\partial L}{\partial y}\right]_y^+ \leq -(o - o^*)^T\frac{\partial L}{\partial o} +$$

$$(y - y^*)^T\frac{\partial L}{\partial y} \leq [L(o^*, y) - L(o^*, y^*)] + [L(o^*, y^*) - L(o, y^*)] \leq 0$$

which indicates that $U_{z^*}(z(t))$ is bounded when $t \geq 0$. According to the Lasalle's invariance principal, the trajectory $z(t)$ asymptotically converges to the optimal point $z_0$, which is the optimal solution of (10) where variables are optimal for (3) and (4).

### B. Proof of Theorem 2

To prove the algorithm is robust to inaccurate damping coefficient $D_i$, following changes are made:

1) $d_i = (C'_{i,e})^{-1}(\lambda_i + \mu_i)$, and $q_i = (C'_{i,h})^{-1}(\delta^+ + \delta^-)$
2) $\gamma^+ \equiv \gamma^- \equiv \zeta \equiv 0$.

The problems (3), (4), and (10) still holds except (9):

$$\dot{\mu}_i = \varepsilon_{\mu_i}[P_i^{in} - d_i + \tau\alpha_i\omega_i - \sum_{j:ij \in \varepsilon_{in}}B_{ij}(\varphi_i - \varphi_j) + \sum_{k:ki \in \varepsilon_{in}}B_{ki}(\varphi_k - \varphi_i)]$$

It can be proved that the time derivative of (12) is bounded:
$\dot{U}(w) \leq \int_0^1 (w - w^*)^T H(w(s))(w - w^*)$ where $w(s) = w^* + s(w - w^*)$. The matrix $H$ is the differential of $\hat{L}$ where

$$H = \begin{bmatrix} (\varphi, Q^v) & (P, \mu_L) & (\mu_g, \lambda_g) & \delta & (\varepsilon, \beta) \\ 0 & 0 & 0 & 0 & 0 \\ 0 & H_{P,\mu_L} & 0 & 0 & 0 \\ 0 & 0 & H_{\mu_g, \lambda_g} & 0 & 0 \\ 0 & 0 & 0 & -q' & 0 \\ 0 & 0 & 0 & 0 & 0 \end{bmatrix} \begin{matrix} (\varphi, Q^v) \\ (P, \mu_L) \\ (\mu_g, \lambda_g) \\ \delta \\ (\varepsilon, \alpha) \end{matrix}$$

where $H_{P,\mu_L}$ and $H_{\mu_g,\lambda_g}$ are both negative semi-definite sub-matrices.

Under assumptions in Theorem 2, since $H \leq 0$, adopting an invariance principle can we prove the robustness for the inaccurate coefficient $\tilde{D}_i$, which is very similar to that in [12].

## V. CASE STUDIES

Case studies research the necessity of considering electric-heat coupling in IES frequency control and the robustness under inaccurate coefficients, which demonstrate the effectiveness of the proposed algorithm.

The topology of the network is shown in Fig.3 [11], in which the conditions are power step change $P_3^{in} = 0.3$ p.u., $Q_3^{in} = 0.3$ p.u., and the heat inertia $Q_3^v = 0.1$ p.u.. The left boundary of the CHP unit at bus 3 is $q_3 \leq 0.5d_3$.

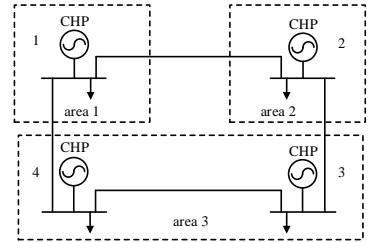

Figure 3. Network topology of case studies

### A. Comparison of Considering/Ignoring Energy Coupling

This case is designed to demonstrate the importance of considering electric-heat coupling in frequency control. *e1* indicates electric-heat coupling is not considered, and *e2* implies electric-heat coupling is considered.

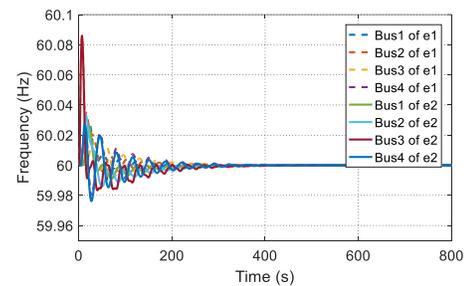

Figure 4. Frequency response to step electric and heat power changes

From Fig. 4 and Fig.5, the frequency can be restored in *e1* and *e2*, but the power outputs are different. Actually, power outputs of *e1* are inaccurate because the constraints of CHP units (6d)-(6e) have been broken, which are reflected accurately by *e2* .

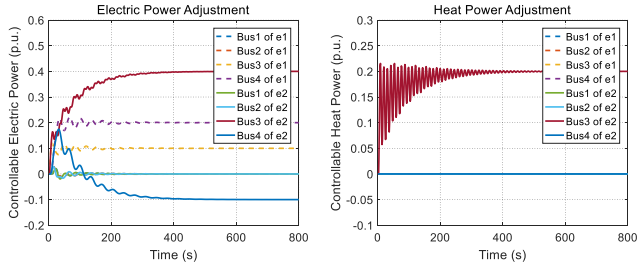

Figure 5.  Power adjustment under step electric and heat power changes

*B. Robustness for Inaccurate Coefficient*

In this case, we study the robustness of the proposed method under a step change $P_3^{in} = 0.3$ p.u. with inaccurate coefficient $\tilde{D}_i$ which is $k$ times of the real damping coefficient $D_i$ i.e. $\tilde{D}_i = kD_i$ .

Shown in Fig. 6, when $k$ is approaching 0, the frequency damping is increasing and the convergence speed is decreasing. But if $k$ is too large, the system becomes unstable and crashes. Under given conditions, the proposed method is robust when $\tilde{D}_i$ varies from 0.1 to 10 times of real value $D_i$ .

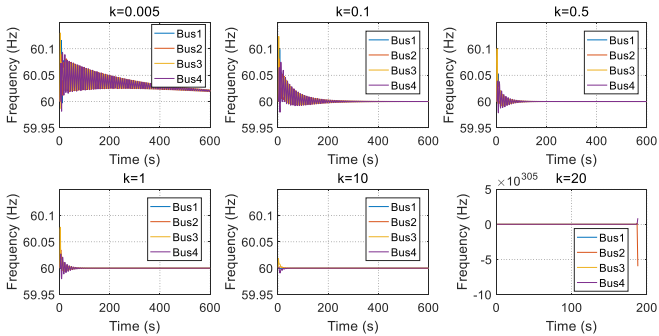

Figure 6.  Frequency response to step electric and heat power changes

## VI. CONCLUSION

We have proposed a fully-distributed frequency control method with system-wide optimality guaranteed in the IES considering electric-heat coupling, line flow limits and inaccurate coefficients. The algorithm results from reverse engineering and only needs local measurement and communication. Case studies show that our proposed method can eliminate frequency deviations with economic optimality even under inaccurate coefficients in the IES, and the electric-heat coupling should be considered to ensure the constraints of CHP units are satisfied.

The future work includes extending our method from transmission level to distribution level, considering heat network, and improving the dynamic performance of algorithm.

ACKNOWLEDGMENT

This work is supported by the Science, Technology and Innovation Commission of Shenzhen Municipality (No. JCYJ20170411152331932) and the China Postdoctoral Science Foundation (2018T110097).